\documentclass[nofootinbib,12pt,aps]{article}
\usepackage{amsfonts}
\usepackage[dvips]{graphicx}
\usepackage{epsfig}
\usepackage[margin=1.15in, paperwidth=8.5in, paperheight=11.0in]{geometry}
\usepackage{tikz}

\def\ben{\begin{equation}}
\def\een{\end{equation}}
\def\bea{\begin{eqnarray}} 
\def\eea{\end{eqnarray}}

\input amssym.def
\input amssym.tex

\newcommand{\SU}{\mathrm{SU}}

\newcommand{\be}{\begin{equation}}
\newcommand{\ee}{\end{equation}}
\newcommand{\bes}{\begin{eqnarray}}
\newcommand{\ees}{\end{eqnarray}}

\renewcommand{\hat}{\widehat}

\begin{document}
\title{\Large On the depth of {\it quantum space}}
\author{\vspace{1cm} \\ Daniele Oriti \\ Max Planck Institute for Gravitational Physics (Albert Einstein Institute) \\ Am M\"uhlenberg 1, D-14476 Golm, Germany, EU}
\date{}
\maketitle


We focus on the question: {\lq\lq Is space fundamentally discrete or continuous?\rq} in the context of current quantum gravity research. In particular, we paint a scenario based on the idea that {\it quantum space} is a sort of peculiar condensed matter system, and on the speculation that its microscopic dynamics is described by a {\it group field theory} formalism. We suggest that, from this perspective, on the one hand the question has no absolute meaning, so no answer, but also that, on the other hand, the reason why this is the case is the {\it quantum space} is much richer and more interesting than we may have assumed. We also speculate on further physical implications of the suggested scenario.

\vspace{1cm}

\section*{Focus question}
\noindent Is reality digital or analog? Of course this question refers, at least implicitly, to the 'ultimate' nature of reality, the fundamental layer. I do not know what this could mean, nor I am at ease with thinking in these terms. 
Therefore, the closest I can get to the issue of the digital/analog nature of reality is to consider a special physical system that is, in some sense, the most fundamental of all: space.


\noindent The question I will focus on in this essay, then, will be: is 
space digital or analog? 
What is the best language to understand/describe (quantum) space? 

\noindent I offer a tentative perspective on this issue, making use of some of the ideas and mathematical models that form the basis of my own work as a quantum gravity theorist.
Also, I will translate the above question into a slightly different one: 

{\bf is (quantum) space discrete or continuous?} 

\noindent Here is my goal: I will phrase the question within a {\it speculative scenario} for quantum gravity, one that is based on an {\it analogy} of space with a condensed matter system, and try convince on this basis that the question makes no (absolute) sense. 

\noindent This may seem no solid basis for an argument, but I assure the reader that this is not going to be (at least, this is not the intention) a Chewbacca defence \cite{chewbacca}. The aim is to make sense of the question in the context of a current quantum gravity model, and to open up a richer perspective on it and on the nature of space itself.

Our current understanding of physical space is encodedin General Relativity (GR). Here space is a continuous system, modeled by a smooth metric 3d manifold, deeply intertwined with time, to form a smooth metric 4d manifold, of given topology and minkowskian signature. The geometry of space is dynamical, and this dynamical nature turns space into a physical system on its own, thus subject to further scrutiny. We also know, from a variety of physical arguments, that this is not the end of the story \cite{QGrev}. 

What can a \lq discrete\rq   space mean? As soon as we leave the above description of space, 
it is not obvious that we can still speak of the physical system as being {\it space} anymore. For example, replacing the smooth manifold structure with some discrete substratum, we loose at once all the defining properties (dimension, signature, topology) 
nor it is guaranteed that they will turn into their continuum analogue in any approximation.

The question becomes: can we define a physical system, that has no feature of {\it space} in itself, such that, in some appropriate (dynamical) regime, it will be described by a {\it space}, \underline{identified}, now, with a smooth manifold? If not, we will not have a more fundamental notion of the physical system we call space, and the discrete/continuum issue will then be settled. If we do, then we will be forced to pose again the question in the context of this new, more general description of this new, more fundamental system, that \lq looks like\rq space in some limited regime. This is why I think it is necessary to pose the question in the context of a specific, if tentative, model of what I will call (for simplicity) {\it quantum space}, and of a specific, if speculative, interpretational framework. 

\

Before we start, let me add a few cautionary remarks.



One risk that I will try to avoid  is the confusion between physical distinction of different regimes of a physical system and of its dynamics, and the  philosophical distinction between different levels of reality, or different ontologies. In particular, the question of the discrete vs continuous nature of quantum space will be addressed in the first context, rather than in the second. Similarly, the problem of the {\it emergence} of a set of properties of quantum space, at one level of description, from a different (more fundamental?) one will be sketched in its possible physical/scientific aspects, in the specific models of quantum space I favor, leaving the philosophical reflection on the respective role of reductionism and emergentism \cite{RedEmer} to a later stage.

The issue of digital/analog nature of Nature, even in the discrete/continuous translation, as well as the related issue of emergentism vs reductionism, are very complicated on their own, in any specific chosen context \cite{RedEmer}. Space is itself a very difficult subject of reflections, for its elusive and at the same time ubiquitous nature. Quantum gravity is an even weirder case study, because it already poses numerous and very difficult philosophical challenges on its own, in part coming directly from its supposed ingredients: General relativity and Quantum Mechanics  \cite{QGPhil,PhilQM,PhilST}. I will not address these challenges, including any issue related to the interpretation of quantum theory when applied to space itself. 


More practically, I will discuss the issue of the nature of space without considering matter (fermions or gauge interactions) in the picture. This is for simplicity only, but could be questioned \cite{RelSubSpace}. 

I will not touch the thorny issue of the emergence of time or of the discrete or continuous nature of the same. In any quantum gravity context, the role of time is subtle to say the least \cite{EmergTime}, and will bring further complications. In particular, I will argue under the assumption that the issue of discrete/continuous nature of space can be disentangled from the same issue about time. Also, I will leave aside the issue of the emergence of time from a regime in which there is no notion of space at all. This absence could be an additional difficulty to our fundamental understanding of time \cite{EmergTime, CarloTime} or a way out of the traditional difficulties \cite{FotiniTime}

In fact, the models of quantum gravity I deal with force us to think in the context of \lq no space\rq , of absence of space, and the issue I will discuss is the extent to which, within these models, one can recover a notion of space at all, and then whether \lq the stuff it emerges from\rq is best understood as  discrete or continuous. 


Finally, I will be taking a \lq realist\rq standpoint concerning the peculiar physical system that is quantum space. However, this will be done for greater agility of exposition, but will not necessarily imply any endorsement of it from the philosophical side. 

For all of the above, it is clear that very many issues will be left aside, and left for future elaboration. I apologize for this.

\newpage

\section*{Quantum gravity perspectives on quantum space}
There is no necessity, coming from either General Relativity or quantum mechanics alone, to deny the smooth structure of space on which they are based. Even curvature singularities in GR are not, per se, an indication that the smooth structure has be disposed of. And quantum mechanics rests on this undisturbed smooth structure in all its successful applications. There are however independent reasons why the classical GR description of space, neglecting quantum properties of the same, could not suffice \cite{QGrev}.

Several arguments, then, indicate that the application of quantum mechanics to space itself may lead to the mathematical or physical impossibility of a point-like, thus continuous substratum for space and geometry: the possibility of forming microscopic black hole in the attempt to resolve points using high energy probes, the necessity of quantum fluctuations in the metric structure, and thus in the determination of distances and causal relations between points, and so on \cite{QGrev,QGPhil}. 

An emergent, approximate validity of a continuous description of space (as in GR) is, however, the natural starting point of any radical approach to quantum gravity \cite{CS,QI,majid} that aims at explaining the {\it origin} of space, its (more) fundamental  {\it structure}, from something that, by definition, then, cannot be described in terms of continuous space and geometry. The one perspective that does not share this attitude is the one that sees the problem of quantum gravity as the problem of {\it quantizing} General Relativity. This is the case, for example, of loop quantum gravity (LQG), including (at least in spirit) its covariant formulation in terms of spin foam models (see \cite{LQG} for extensive treatments). 
However surprises come, as often, from unexpected corners. 
Even within the canonical approach, an immediate hint that some discrete structure of space results simply from the quantization process is the discrete spectrum of (kinematical) geometric observables like areas and volumes. 
This is the analogue of the discretization following quantization of, say, the energy spectrum of the hydrogen atom. 
Even more important, in LQG, a generic wave function of the quantum gravitational field turns out to be a superposition of excitations of the same living on {\it graphs}, thus purely discrete structures. 
So to speak, there is no {\it space} outside these graphs and the data labeling them. Thus purely discrete structures are suggested as the true substratum of {\it quantum space} even in approaches that aim at simply bringing together in full General relativity and Quantum Mechanics. 
A similar discourse can be made for simplicial quantum gravity approaches \cite{simplicialQG}.
These start as straightforward implementations of the quantum gravitational path integral, by means of a regularization of both geometry and GR dynamics using simplicial lattices. 
Still, in practice one finds oneself with a quantum theory of discrete structures, from which a continuum space with continuum GR dynamics should only emerge in a limit in which an infinity of these discrete structures interact. 
Thus, continuum space ends up being once more an {\it emergent} construction from more fundamental structures that are not continuum, not space \cite{simplicialQG}. 

Thus we learn two lessons: 1) the disposal of the current continuum definition of space in favor of {\it something different}, possibly discrete, can be the natural outcome of straightforward joining of Quantum mechanics and General Relativity; 2) the very same structures that one arrives at in doing so admit a more radical re-interpretation than initially devised.  
Group field theories use the same discrete structures of LQG and simplicial quantum gravity, suggesting a radical re-interpretation of them, and, we will argue, a vastly richer perspective on the discrete/continuum nature of quantum space.

\section*{The basic analogy: condensed matter systems}
The analogy that sheds a new light on the question of the discrete/continuous nature of {\it quantum space} is to see it as a condensed matter system, a weird \lq material medium\rq. 
This radical idea is not new.
It has been advocated in \cite{hu}, the similarities of GR with elasticity theory in solids have been described in \cite{paddy}, the possibility of deriving the Einstein's equations for the dynamics of space as an equation of state for some underlying microscopic system has been studied in \cite{GRthermo}, the unreasonable attitude of treating space as a \lq\lq non-substance with substance-like properties\rq\rq has been criticized in \cite{laughlin}, several studies of analogue gravity systems in condensed matter show that some properties of our space emerge naturally in a variety of \lq less fundamental\rq systems \cite{analog}.
This set of results suggests that space itself can be understood as a kind of Aether (a medium-substance) that is, however, relativistic (at least up to Planckian scales), whose structure has to be described in background independent manner, and is fundamentally quantum in nature. The physical question, then, becomes what constitutes it, what it looks like at microscopic scales. 
The novelty is now that this idea has the possibility of being realized rigorously, within a fastly developing quantum gravity approach: group field theory. 


 Suppose I give you a condensed matter system, the assembly of a huge number of given building blocks of some type, somehow holding together: a bucket of superfluid Helium, or simply a tank of water. Now I ask: is it discrete or continuous? 

The first point to notice is that this is a physical question. It amounts to ask whether the best mathematical description of the system, that is able of reproducing or predicting its physical properties, is given in terms of discrete or continuous entities. 
Second,  the true answer to the above question, when posed in these terms, is: it depends. 
Whether the physical system considered is best described as discrete or continuous depends on two set of things: what is the range of physical macroscopic, thermodynamic parameters at which we are considering the system, and the observation scale we adopt. The first aspect has to do with the {\it phase} in which the system is, and we know well that its physical properties are going to be very different in different phases, including its being well described as a continuous or a discrete medium. The second has to do with the fact that, even within a given phase, the {\it observation scale} one uses affects which properties of the system are relevant and which negligible, and what type of description is best suited to adopt them. 

Think of water in a bucket. Is it discrete or continuous? First of all I have to pose the question within a certain thermodynamic phase. At higher temperature, say 600K, and given pressure, say $10^6$ Pascals, the substance we call water is a gas, which can be described by a fluid with very low density and high kinetic energy, a continuous system possibly  homogeneous and isotropic. At lower temperature, say around 350K, and same pressure, we find again a continuous fluid, a liquid (the thing we normally call water), with much lower kinetic energy, higher density and very different physical properties (interaction with matter or light, etc), possibly still homogenous and isotropic. The best description of it will be given by a continuum field theory, hydrodynamics, for the density and the velocity of the whole fluid. At even lower temperature, say below 200K, the same system will look and behave very different; it will be a solid, a crystal usually called \lq ice\rq, which is rigid, does not flow like water, and has lost the continuum symmetries. Its best description will then be that of a {\it discrete} structure, an hexagonal lattice, and the interaction with other systems will now be different. So, is the system we call water discrete or continuous? In a very strict sense, what we call water is the liquid, continuous phase of the system, so it is continuous by definition, as we suggested for physical space. In a broader  sense, it depends. 
It depends on the phase in which the system resides, which in turns depends on the macroscopic parameters characterizing it. Obviously, what we said for water holds true for any condensed matter system, say a quantum fluid like a Bose condensate, and the main problem of condensed matter theory is, given a certain physical system, to identify and characterize the various phases in which it can be found, the physical properties of each phase, and the transitions between phases, by methods of statistical field theory, e.g. the renormalization group or mean field theory.  

The same could be true for {\it quantum space}.  

Let me counter-argue a possible objection to the above argument: \lq\lq Sure, at macroscopic scales a condensed matter system like water or a Bose fluid may look continuous, at least in some phases, and the discrete/continuum dichotomy is not a fundamental one, but something dependent on the phase one finds the system in; however, {\it fundamentally} the system is discrete, because it is made up of molecules or atoms, which are the fundamental, quantum building blocks of it, and the macroscopic phases are just different ways in which these fundamental building blocks organize themselves\rq\rq. 

In particular, one may point out the aspect of observation scale. At smaller scales or higher energies, we know that the hydrodynamic, continuum description of a fluid fails. A different type of description is needed, in terms of the microscopic constituents, atoms or molecules. 
Indeed, it is a fact, although a non-trivial one (it is not easy to specify clearly what we mean by \lq made of\rq, by \lq just organize themselves\rq etc), and I tend not to quarrel with facts. Still, I do not think it changes the answer to the question; it leads to refining it further. 
I have two problems with the idea that a condensed matter system is fundamentally discrete because it is made of atoms. The first is conceptual. This point of view rests on a certain reductionism that I accept, like most physicists \cite{Anderson, everything}, but still find in need for further analysis. More important, it also seems to rest on a constructivist ideology, i.e. on the assumption that, at least as a matter of principle, it would be possible to reconstruct or derive {\it all} properties of the macroscopic system, the collective behaviour of large numbers of microscopic building blocks, from a good enough description of them. This is really the point of view that the \lq emergentist\rq paradigm \cite{RedEmer} opposes. My point here is that this is more of a philosophical issue, rather than a physical one, and a subtle one as well, so our scientific answer cannot rely on it, in the even subtler case of quantum space. The second, more important problem I have is with the idea that the microscopic building blocks of a condensed matter systems are themselves fundamentally discrete. It seems to me that also this is not absolutely true, that \lq it depends\rq. 
This is simply because, before any further approximations, of limited validity by definition, the framework for describing the microscopic properties and dynamics of the very building blocks of any condensed matter system is quantum field theory, in which the atoms are quanta of a continuous field. 
One can speak of individual quanta within a specific presentation of the field theory, the Fock representation, and, from the physical point of view, when only a finite (possible small) number of excitations around the Fock vacuum (atoms) are relevant. The question whether the system, at microscopic scales, is discrete or continuous boils down once more to a matter of convenience of a certain physical description over another, equally fundamental. It also becomes tied to the physical  question of whether the regime in which only a limited number of atoms exists, i.e. only a few quanta of the field are excited, is experimentally accessible, or to the other physical question of whether some specific property of the system as a whole can be traced back directly to the atoms. 
And again, even this description was relevant, it would be so only in some specific physical situation, to capture some specific features of the physical system.

So, is the system discrete or continuous? 

It depends.

\newpage

\section*{A candidate theory of atoms of quantum space}
The idea I am arguing for is that of {\it quantum space} as a kind of condensate \cite{GFTfluid}, made of \lq fundamental\rq building blocks that cannot be understood, in themselves, in terms of the ordinary notions of space or time. Moreover, I am arguing that, from this perspective, the question whether {\it quantum} space is discrete or continuous makes no absolute sense, i.e. it has to be answered differently in different physical contexts, as for any other condensed matter system. This last point, as shown, rests on the fact that even the microscopic building blocks of a condensed matter system  are best described in the context of quantum field theory. 

In fact, 
we do have a candidate QFT formalism for the microscopic structure of {\it quantum space}: group field theory (GFT) \cite{iogft}
I will now sketch its basic features.

We could motivate the GFT framework from several other approaches: matrix models, simplicial gravity, loop quantum gravity \cite{iogft}. 
LQG \cite{LQG} starts from the reformulation of GR in terms of $\SU(2)$ connection variables and conjugate triad fields in the continuum. Still, one ends up with a Hilbert space decomposed into sums of cylindrical functions which only depend on a finite number of holonomies (group elements $g$), associated to a given graph
For the same quantum states, one can define two other representations: a non-commutative flux representation \cite{ioaristide}, in which the links of the graph are labelled by Lie algebra elements $x$ of $\SU(2)$, representing triads,  and one in which links are labelled by $\SU(2)$ representations $j$, quantum numbers of geometric operators. 

Now imagine decomposing the (say, trivalent) graphs into their constituting (open) {\it vertices} with corresponding re-arranging of labels, so that instead of a graph we now have a collection of open vertices with some \lq gluing\rq relations defining the original graph.

\begin{minipage}[l]{5cm}
\includegraphics[width=4.8cm, height=3.8cm]{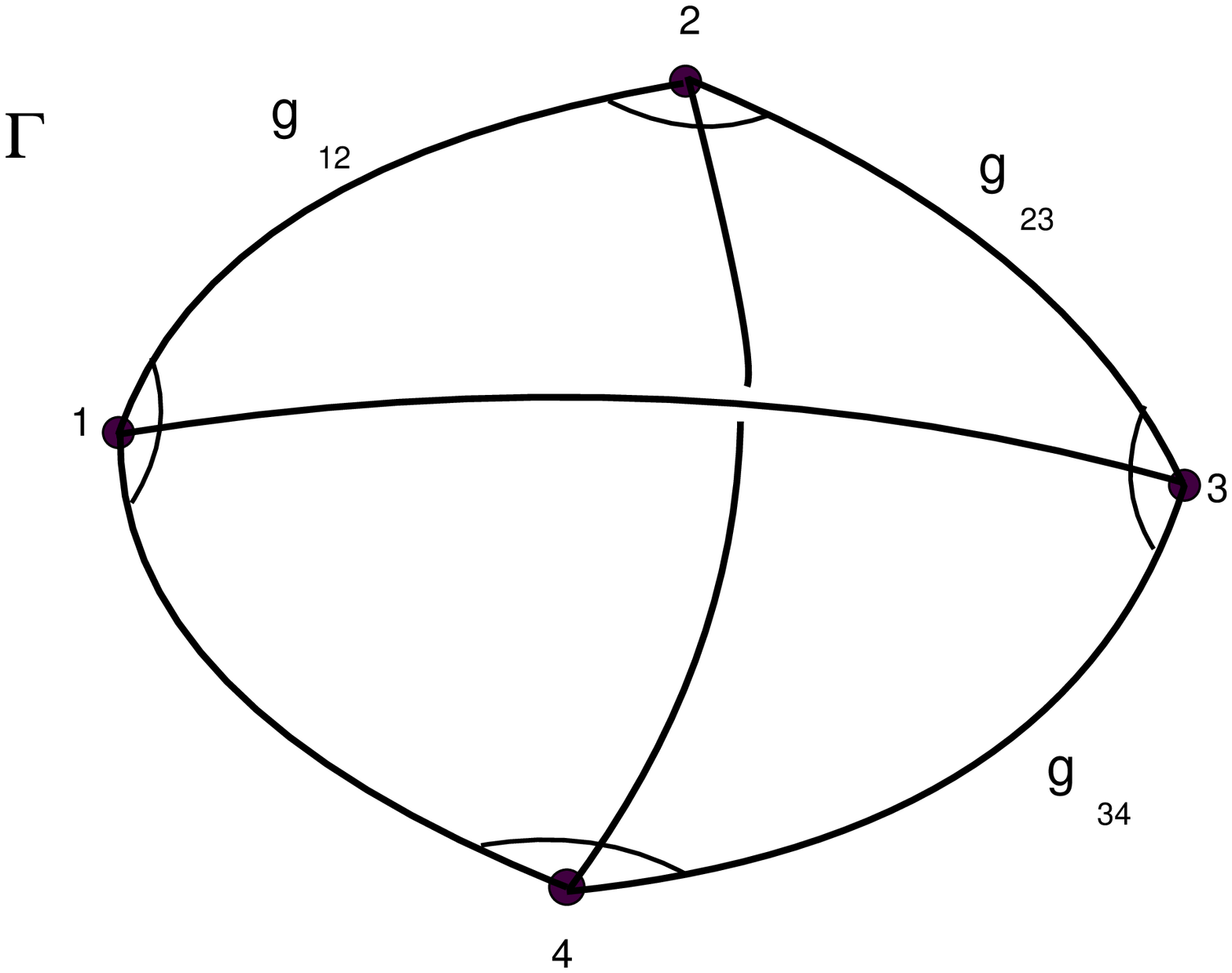}
\end{minipage}
\hspace{0.5cm}
\begin{minipage}[l]{3cm}
\includegraphics[width=4.8cm, height=3.8cm]{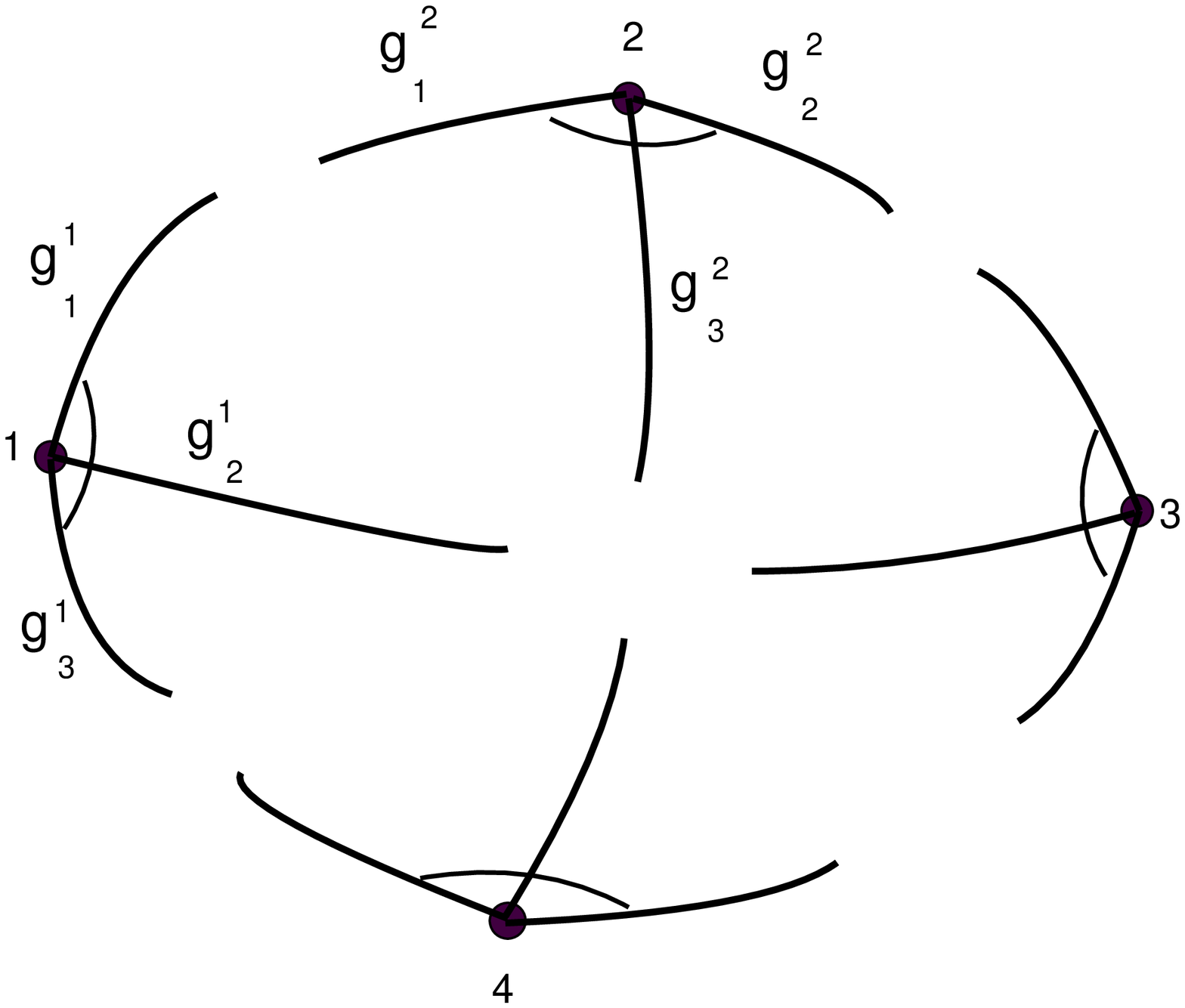}
\end{minipage}

These are the fundamental LQG building blocks of {\it quantum space}: combinatorial structures with associated fundamental degrees of freedom being purely algebraic. 

Simplicial quantum gravity comes into play because the same vertices can be understood as combinatorially dual to triangles with the same associated degrees of freedom. 
The corresponding wave functions for arbitrary graphs would then be interpreted as wave functions associated to a simplicial 2d space in a discrete quantum gravity approach based on connection variables. 
For individual vertices one would then have wave functions and graphical representation as:

\includegraphics[width=14.4cm, height=4cm]{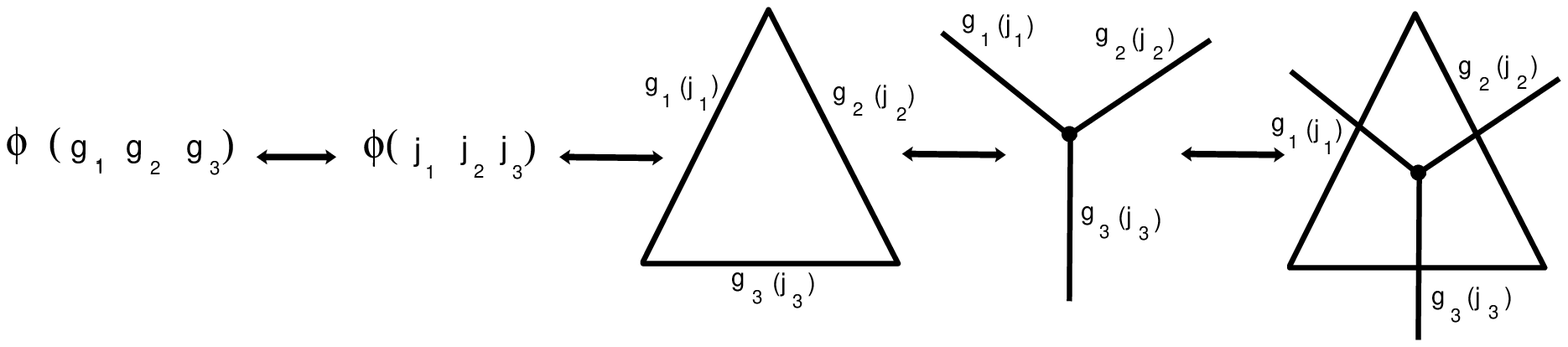}

This means that the same wave function can be understood as defining a possible state of a quantized simplex, i.e. a chunk of quantum space, in a discrete approach to quantum gravity \cite{SF, iogft}. 
A similar reinterpretation is available in higher dimensions \cite{LQG,iogft}.

Now, take these simple structures seriously as the definition of the building blocks, the atoms, of {\it quantum space}. 
A generic configuration of {\it quantum space} would be given by an arbitrary number of them. Consider the case in which such number is not conserved, but they can be created/annihilated (that is, the combinatorial structure of a generic configuration of {\it quantum space} varies) or top of changing their associated labels (in the same sense in which atoms can vary their position, energy, etc). The natural dynamical framework would then be that of a quantum field theory, in which the single-vertex wave function becomes a quantum field. This is a GFT \cite{iogft}.


As in LQG, we have three equivalent representations for the basic field: with group variables, with Lie algebra variables and with representation variables \cite{LQG,iogft,ioaristide}:

 $$ \varphi(g_1,g_2,g_3) \leftrightarrow \varphi(x_1, x_2, x_3) \leftrightarrow \varphi^{j_1j_2j_3}_{m_1m_2m_3} $$
 
 The interpretation of the $x$ variables is that of elementary edge vectors associated to the edges of the quantized triangle, that of the $g$ variables is of parallel transports of the gravity connection along links of the dual graph vertex, and that of the representation parameters $j,m$ is that of quantum numbers of geometry. 

The convolution of multiple fields represents the gluing of triangles or graph vertices along common edges or links, and thus the formation of more complex discrete structures, representing extended configurations of {\it quantum space}.
 
\begin{figure}[here]
\includegraphics[width=14.1cm, height=3.1cm]{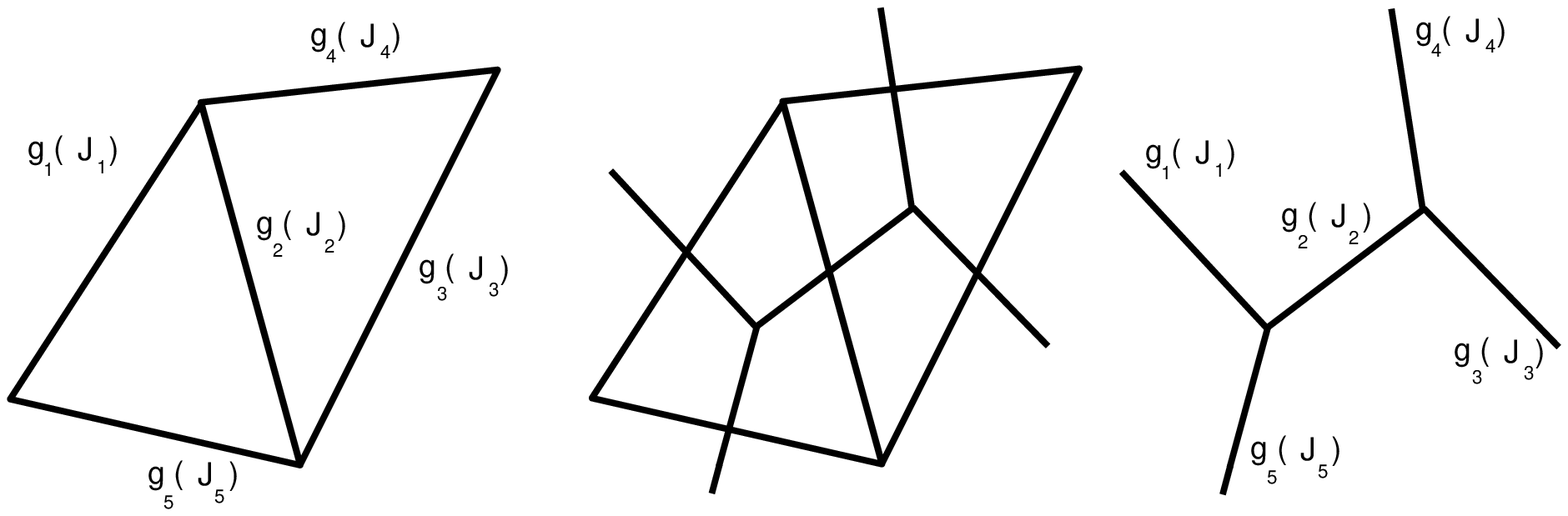}
\end{figure}

The GFT dynamics is dictated by very minimal considerations. Noting $\varphi_{123}:=\varphi(x_1,x_2,x_3)$ we define, schematically, the action \cite{iogft, ioaristide}
\ben
S =\! \frac{1}{2}\int  \,\varphi_{123}\, \varphi_{123} -  
\frac{\lambda}{4!} \! \int  \,\varphi_{123} \, \varphi_{345}\,  \varphi_{526} \, \varphi_{641}
\label{gftaction}
\een

In the interaction, four triangles are glued to form a tetrahedron by edge identification. The kinetic term glues two tetrahedra along a common triangle. 

The quantum theory is then defined by the perturbative expansion of the partition function in powers of the coupling constant $\lambda$. 
$$ Z\,=\,\int
\mathcal{D}\varphi\,e^{-S[\varphi]}\,=\,\sum_{\Gamma}\,\frac{\lambda^N}{{\rm sym}[\Gamma]}\,Z(\Gamma),
$$
where $N$ is the number of interaction vertices in the Feynman diagram
$\Gamma$, ${\rm sym}[\Gamma]$ is a symmetry factor for $\Gamma$ and
$Z(\Gamma)$ the corresponding Feynman amplitude.

By the combinatorial structure of vertices and propagators, the Feynman diagrams will be dual to 3d simplicial complexes, i.e. a primitive, discrete version of a \lq spacetime\rq. Looking at the Feynman amplitudes one sees the link with simplicial quantum gravity and loop quantum gravity.
In Lie algebra variables, the amplitude $Z(\Gamma)$ for a generic Feynman diagram is exactly \cite{ioaristide}
the simplicial path integral of 3d gravity in 1st order form.
The expression of $Z(\Gamma)$ in terms of group representations, in turn, 
is the Ponzano-Regge spin foam model \cite{LQG}, also defining the dynamics of 3d loop quantum gravity. 

The above construction can be straightforwardly extended to give transition amplitudes between arbitrary configurations of {\it quantum space}.
A discrete spacetime substratum, then, emerges as a possible interaction process of GFT quanta. The dynamics is given by a sum over all possible discrete topologies, each weighted by a discrete, algebraic version of the gravity path integral.  

Notice that, while several ingredients from classical GR enter the construction, at least as a motivation, the GFT dynamics is {\it not} the result of any straightforward quantization of GR dynamics.
Rather, if the GR dynamics is associated, in the analogy, with the hydrodynamics of {\it quantum space}, the spacetime equivalent of \ref{hydro}, the GFT dynamics has to be compared with the microscopic atomic theory of the same system \ref{TheoryOfEverything}.

So, in GFT we have a class of models of {\it quantum space}, possibly fundamental, that do not share language and structures used in GR description of space. Indeed the main problem becomes to recover  this in some limit.
How to do so?

We most likely need a very large number of GFT quanta to constitute an effectively continuum region of
space, governed by continuum gravitational
physics. This is probably governed by collective dynamical laws, not by the microscopic GFT dynamics (the analogue of \ref{TheoryOfEverything}). Finally, whatever the exact details of relevant phase is, we expect our condensed matter system,
i.e. {\it quantum space}, to be close to equilibrium. 
This seems just a description of a peculiar quantum fluid, governed by hydrodynamical equations (the analogue of \ref{EulerEquation}).
Notice that this continuum approximation is conceptually independent of a semi-classical approximation. The emergence of continuum space may even be {\it the result of a purely quantum property of the system}. 

The picture is then the one we had anticipated:
{\it quantum spacetime as a (quantum) fluid of GFT particles, governed microscopically by the GFT partition function, but macroscopically by some GFT effective
hydrodynamics.} \cite{GFTfluid,GFThydro}

This is at present just a speculation. It immediately implies one thing, however: 
to recover and describe a continuum space and its dynamics, it is convenient, if not necessary, to go beyond the spin foam or simplicial gravity description of
the GFT system, since the perturbative GFT formulation, in which they appear, is mainly useful for describing the system in its few-particle regime. For collective behaviour, we should move away from the no-particle vacuum.

Instead, 
we need to use a {\it statistical group field theory} formalism, e.g. renormalization group ideas, for identifying the different phases of the theory, hoping that some GFT models lead to the existence of at least one
with the right properties for a continuum geometric description. This is what the analogy with condensed matter systems suggests for {\it quantum space}. Second, we need to develop first and then use an {\bf
{\it effective field theory or hydrodynamic description}}, coming from the fundamental GFT, for describing the dynamics in the different phases, using again tools from condensed matter theory, e.g. mean field theory. Work in all these directions, partly motivated by the above considerations, is well under way \cite{iogft,GFThydro,GFTrenorm}.
I do not intend to review it here. 


My goal here is only to highlight, within this speculation, how our picture of space would change, concerning the initial question: is (quantum) space discrete or continuous?


\newpage

\section*{Discussion}

We now discuss in some more detail the physical aspects of the scenario for {\it quantum space} we sketched, based on the GFT formalism.  

Is it the return of the Aether? In some sense, it is, as we are arguing that there exists a sort of weird material medium, filling everything (better, \lq constituting everything\rq), having its own dynamics and interacting with matter in a way that (at the very least) influences their respective dynamics. But this is just, in our intention, the appropriate reading of: the very content of GR as a theory of space(time) endowing it and its geometry with the characteristics of a dynamical, physical system; 2) the disconnected, but intriguingly converging results mentioned above suggesting the possibility of interpreting in thermodynamic or hydrodynamical terms the same equations of GR; 3) the current tentative picture and developed mathematical structures arrived at by several approaches to quantum gravity. If it is a sort of Aether, that we are talking about, it is a new type of it, a background independent, pre-spacetime kind of condensed matter system. We are not moving backward from the revolutions of Relativity and Quantum Mechanics, on the contrary we are trying to move forward on the basis of them. 

The picture we painted shares a certain spirit with the ideas of gravity as thermo- or hydrodynamics, and even more with other background independent approaches to quantum gravity, also inspired by condensed matter ideas, e.g. the {\it quantum graphity} programme \cite{QI}, but it framed in a mathematical framework that is the direct generalization of matrix models of random surfaces \cite{mm}. In particular, our proposed scheme for the emergence of continuum geometry from the dynamics of the GFT quanta, as a phase transition, can be seen as a sort of {\it geometrogenesis} \cite{QI}.

One is naturally led to further speculations about the true meaning of spacetime singularities and the corresponding breakdown of the GR description of space. In an approach to quantum gravity guided by the idea of \lq quantizing GR\rq only, this breakdown signals the need to take into account quantum effects of geometry, and, a priori, nothing more. That is, the breakdown of the semi-classical approximation in the description of a quantum, continuum space. 
Being more radical, 
it may also signal the breakdown of the continuum idealization of quantum space. In the condensed matter analogy, this would mean the breakdown of the hydrodynamic description, classical or quantum, of the same, and the need to resort to a more fundamental description of it in terms of its building blocks
. In the same analogy, one could put also forward the hypothesis that something even more radical is at play in singularities: a phase transition of quantum space, from a condensed, fluid phase to something completely different, where most of the defining notions of space to fail to apply: locality, and thus local spacetime symmetries, the notion of a fixed dimension, etc.
The approach to the transition, could be maybe modeled by more standard looking, but exotic effective models defined on a continuum space, e.g. field theories with deformed relativistic symmetries \cite{DSR} or varying speed of light \cite{vsl}, or living on fractals \cite{fractal}. 
The idea of a cosmological phase transition {\it of space itself}, replacing the Big Bang singularity, may provide a novel way to look at the puzzles of very early cosmology (horizon problem, flatness problem, etc), currently address by inflation, itself in need for a better explanation. 


Also the \lq emergence of space\rq is not to be understood as happening \lq in time\rq, unless: a) it happens after, mathematically speaking, some degrees of freedom of the system decouple to define a {\it pre-geometric notion of time}, e.g. a thermodynamical notion of time, or 
b) some of the macroscopic (order) parameters that define the phase diagram of the GFT system turns out to have the interpretation, at least in the geometric phase of the GFT system, of a geometric time. 

Now we come to a rather thorny issue.
This is what Laughlin calls {\it the Dark Side of Protection} \cite{laughlin}. Is the microscopic model that realizes the \lq emergence of space\rq irrelevant at least to a large extent?
The point is that a number of macroscopic features of a condensed matter system are largely independent of the underlying microscopic physics, and are instead {\it universal}, e.g. due only to very general symmetry properties and a choice of basic dynamical variables. 
This is in general a good thing, because it means that we do not need to get the microscopic theory exactly right, whatever that means, in order to understand the macroscopic property of the system. In a sense, this is what makes condensed matter theory possible at all, by \lq protecting\rq the interesting macroscopic properties from the microscopic and usually out of control details of the dynamics of the building blocks. Something similar could be at work for quantum space. 
For example, if the relevant symmetry property of the continuum phase is diffeomorphism invariance and the relevant dynamical variable is the metric field, one could use very efficiently the framework of effective field theory \cite{EFT} for doing quantum gravitational physics even at high energies. That this is possible is of course a good thing. The danger is that this could be the best we can do, even with a complete model of the microscopic details of quantum space. Even if it does not lead to any breakdown of the effective field theory framework, such a model would be expected to help us at least predict the constants of the effective field theory for gravity we use at macroscopic scales. But what if even these turn out to be largely universal, thus independent of the details that distinguish, say, one microscopic model from another? how do we compare different microscopic models, then? and is then of any use to have them at all, if only very general properties of them are falsifiable? 
However, one can see the same issue more optimistically. We have, at present, not a single complete candidate for the microscopic theory of quantum space, in particular not  single one for which we are able to show how it gives rise to a continuum space at macroscopic scales. Universality can be useful in helping us to identify a class of models, characterized by very general features (e.g. symmetries, type of variables) that succeed in doing so (e.g. predict the right effective coupling constants, in an effective field theory formulation of quantum gravity, or the correct critical exponents, etc), without indeed having to \lq get all the details right\rq, and possibly capture them in simplified, effective models. 
Moreover, if the emergence of the correct continuum space is due only to some very general properties of the (GFT) atoms of space and of their dynamics, this also means that it is enough to understand this generic link in order to open up an entirely new perspective on space and on how it comes about. 

However, the physical reality of this atomic level, and the \lq fundamental discreteness\rq of quantum space could be asserted {\it only} if it is accessible to direct or indirect experimental test. This is all the more true if we want to claim the existence of, and philosophize about, an entirely different phase for the macroscopic system that quantum space is, one in which the very notion of continuum space is not applicable. This is far beyond the simple-minded difficulty of accessing, by direct experiments, Planck scale effects, true for any theory of quantum gravity. One indirect test would be to explain some puzzling (large scale) feature of continuum space (e.g. dark energy?) in terms of generic properties of its quantum, atomic structure. Having a theory of {\it quantum space atoms}, and being smart enough, we may be surprised to find out that there are plenty such experimental inputs, even if they do not refer {\it directly} to the atomic properties of {\it quantum space}. As we learned \cite{lakatos}, it is often the theory that guides us towards recognizing new phenomena around us, we cannot blindly search for them simply by increasing the resolution or the energy of our probes into nature. 
Still {\it quantum space} can be declared to be discrete only if: a) we have experimental corroboration for a theory of the (GFT) quanta of space, {\it and} 2) we have access to the \lq few particles\rq regime of the same theory. If we only have access to phenomena depending on the full, field-theoretic, and therefore continuous, description of the same atoms/quanta of space, then we could not sensibly declare that space is fundamentally discrete in any sense. 



We may not achieve the same level of certainty about the nature of {\it quantum space} as we have for quantum matter, but we will have found a more interesting reality than we naively assumed in imagining theories of everything or fundamental quantum theories of gravity in the traditional sense.
We will have to accept and then understand the existence of different levels and phases of reality, and their mutual relations. 

So, is quantum space continuous or discrete? If the above speculation is right, to realize it concretely will be a revolutionary scientific and cultural experience. But it will not provide us with a better answer to this question than: \lq\lq It depends\rq\rq.

\noindent {\it Quantum space}, as the rest of Nature, is much richer and more interesting than we think.

\newpage
\section*{Appendix}
Consider the basic description, both microscopic and macroscopic, of a very interesting condensed matter system, a Bose condensate of Helium-4 atoms \cite{hu}.
The hydrodynamic description of the superfluid phase of the system, after condensation of the atoms has occurred \cite{volovik, cdt}, is based on the density of fluid $\rho(x)$ and the velocity of fluid ${\bf v}(x)$:

\be
\partial_t \rho=\left\{H,\rho\right\}= -\nabla\cdot(\rho{\bf v})~,
\;\;\;\;\;\;\;\;
~\partial_t{\bf
v}=\left\{H,{\bf v}\right\}=-({\bf v}\cdot\nabla){\bf
v}-\nabla\frac{d\epsilon}{d\rho}~.
\label{EulerEquation}
\ee

for some energy density $\epsilon(\rho)$. These equations, for vortex-free flows, and $ \bar\kappa\equiv 2\sqrt{K\rho}
$, ${\bf v}=\bar\kappa\nabla \theta$, $\Psi=\sqrt{\rho}e^{i\theta}$, with constant $K$, can derived from the extended Lagrangian (with gradient terms):

\be \label{hydro}
L_{\rm
GP}(\psi)=  \int d^4x \left(\frac{i\bar\kappa}{2}\left(\psi^*\partial_t
\psi-\psi\partial_t \psi^*\right)
+\frac{\bar\kappa^2}{2}\nabla\psi^*\nabla\psi+
\epsilon(\rho)-\mu\rho\right)~,~\rho=|\psi|^2~.
\ee

which describes a generalized Gross-Pitaevski hydrodynamics.
So we get a classical field theory describing our continuum system, a fluid, at macroscopic scales. In the analogy of continuum space as a condensate, this would be the analogue of GR. As for GR, one would discover that there are properties  of the system at macroscopic scales (e.g. superfluidity) that cannot be fully understood remaining at this level of description, or that the above description itself breaks down in some regime. So one would look for a better theory. As in the gravity case, the first possibility would be to simply \lq quantize\rq the classical theory, turning classical (phase space) variables to operators and so on: 
${\bf v}(v) \rightarrow \hat{\bf v}(x)$, $\rho(x) \rightarrow \hat{\rho}(x)$, $\psi(x) \rightarrow \hat{\psi}(x)$.
The resulting quantum theory is perturbatively non-renormalizable (with quadratic divergences, just like GR), so in general would be only an effective field theory.
However, the resulting quantum theory is not so interesting,  because before quantum fluctuations of $\hat{\rho}$ or $\hat{\bf v}$ become relevant, the whole hydrodynamic approximation breaks down, and the microscopic atomic structure of the fluid becomes relevant. In this case, we also know what is the general form of the correct microscopic theory for the quantum atoms (the theory of everything):
\begin{eqnarray}
L_{\rm micro}= \int d^4x \,\Psi^*\partial_t
\Psi-\Psi\partial_t \Psi^*\,+\,\Psi^\dagger({\bf
x})\left(-\frac{\bar\kappa^2}{2}\nabla^2  -\mu
\right) \Psi({\bf x}) 
\nonumber
\\
+{1\over 2}\int dt d^3x\int d^3y ~\Psi^\dagger({\bf x})
 \Psi^\dagger({\bf y})U({\bf x}-{\bf
y}) \Psi({\bf y}) \Psi({\bf x}),
\label{TheoryOfEverything}
\end{eqnarray}
It is a continuum field theory for the fields whose quanta are the atoms constituting, in the appropriate condensed phase, the quantum superfluid, with a quartic potential representing the pairwise atomic interaction.

Now let's consider the opposite type of problem: assuming we had been given the above correct, quantum description of the atomic system at the most fundamental level, how would we go about extracting the correct behaviour at macroscopic scales? would this be in terms of continuous or discrete entities in its most useful formulation? and how would one derive the hydrodynamic description? 
Well, this system is reasonably well understood (despite continuous surprises and interesting facets, so we know what to do. First of all, we know that the full thermodynamic limit of  the system, in which all of its infinite, interacting degrees of freedom play a role, will be characterized, in general, by a variety of different phases, depending on the temperature, pressure, and other macroscopic parameters, and on (some of) the properties of microscopic building blocks. We also have the right powerful tools to analyze these phases and the phase transitions, i.e. statistical field theory. We know that an hydrodynamic approximation will be valid in one of these phases. We also know, by a variety of experimental inputs, theoretical assumptions and simplified models, general symmetry considerations etc, that in one of these phases, in fact, the atoms will condense and give rise to a collective behaviour characterized by a small number of collective hydrodynamic variables. We start by approximating microscopic dynamics with a simpler one capturing the key properties:

\begin{equation} \label{Ham}
\hat{H} = \int 
\hat{\Psi}^{*}(x) \left( -\frac{\hbar^{2}\nabla^{2}}{2m}  + \frac{\kappa}{2} \hat{\Psi}^{*}(x) \hat{\Psi}(x) \right) \hat{\Psi}(x)
\,d^{3}x.
\end{equation}
Then, we move away from Fock vacuum $\hat{\Psi}\mid F.S.\rangle = 0$ to a non-trivial vacuum state, describing the condensate state such that $\hat{\Psi}(x) | G.S. \rangle \approx \psi(x) | G.S. \rangle $, so that  $\psi$ is our order parameter characterizing the new vacuum (phase) of the system. For example, a state with this property is a (2nd quantized) coherent state:

\begin{equation} |z_{i}\rangle = e^{-|z|^{2}/2} \exp(z_{i} \hat{a}_{i}^{\dagger}) |F.S.\rangle\; .\end{equation}
Then, one assumes that the system is in a macroscopic configuration close to $\mid G.S.\rangle$ and applies a mean field approximation:
\begin{equation}\hat{\Psi}(x) \approx \psi(x) \mathbb{I} + \hat{\chi}(x),
\end{equation}
with $\psi(x)$ the condensate wavefunction, and $\hat{\chi}$ deviations from the mean field $\psi$. Finally, the microscopic equations for $\hat{\Psi}$ give then rise to the effective equations for the order parameter $\psi$, which are the GP hydrodynamic equations.

\

For more technical details on the group field theory formalism, and its relation with other quantum gravity approaches, we refer instead to the literature cited \cite{iogft, GFThydro, GFTfluid, ioaristide}.

\newpage

\end{document}